\pdfoutput=1
\documentclass[
 reprint,
superscriptaddress,
groupedaddress,
nofootinbib,
 amsmath,amssymb,
 aps,
 pra,
]{revtex4-1}
\usepackage{graphicx}
\usepackage{bm}
\usepackage{hyperref}
\usepackage[mathlines]{lineno}

\usepackage[utf8]{inputenc}
\usepackage{amsmath,amssymb}
\usepackage[T1]{fontenc} 
\usepackage{microtype} 
\usepackage[english]{babel} 

\usepackage{booktabs} 

\usepackage{graphicx}
\usepackage{amssymb}
\usepackage{braket,mleftright}
\usepackage{empheq}
\usepackage{subfigure}
\usepackage{stackrel}
\usepackage{soul}
\usepackage{blkarray}
\usepackage{multirow}
\usepackage{amsmath}
\usepackage{physics}
\usepackage{amsfonts}
\usepackage{bm}
\usepackage{bbold} 
\usepackage{color}
\newcommand{\beq}{\begin{equation}}
\newcommand{\eeq}{\end{equation}}
\newcommand{\bse}{\begin{subequations}}
\newcommand{\ese}{\end{subequations}}
\newcommand{\bea}{\begin{eqnarray}}
\newcommand{\eea}{\end{eqnarray}}

\usepackage[utf8]{inputenc} 
\usepackage{amsmath}

\usepackage{float}

\usepackage{enumitem} 
\setlist[itemize]{noitemsep} 
\usepackage{hyperref} 
\usepackage{braket}
\usepackage{verbatim} 

\begin{document}


\title{Many-body localization and the emergence of quantum darwinism}

\author{Nicol\'{a}s Mirkin}
\email[Corresponding author:]{\,mirkin@df.uba.ar}
\affiliation{%
Departamento de F\'{i}sica “J. J. Giambiagi” and IFIBA, FCEyN, Universidad de Buenos Aires, 1428 Buenos Aires, Argentina
}%

\author{Diego A. Wisniacki}

\affiliation{%
Departamento de F\'{i}sica “J. J. Giambiagi” and IFIBA, FCEyN, Universidad de Buenos Aires, 1428 Buenos Aires, Argentina
}%

\date{\today}%

\begin{abstract}
Quantum darwinism (QD) is the process responsible for the proliferation of redundant information in the environment of a quantum system that is being decohered. This enables independent observers to access separate environmental fragments and reach consensus about the system's state. In this work, we study the effect of disorder in the emergence of QD and find that a highly disordered environment is greatly beneficial for it. By introducing the notion of lack of redundancy to quantify objectivity, we show that it behaves analogously to the entanglement entropy (EE) of the environmental eigenstate taken as an initial state. This allows us to estimate the many-body mobility edge by means of our darwinistic measure, implicating the existence of a critical degree of disorder beyond which the degree of objectivity rises the larger the environment is. The latter hints the key role that disorder may play when the environment is of a thermodynamic size. At last, we show that a highly disordered evolution may reduce the spoiling of redundancy in the presence of intra-environment interactions.    
\end{abstract}

\maketitle

\section{Introduction} 
The question of how our classical experience emerges from the quantum nature of reality is a fascinating problem that has been a matter of controversy since the origin of quantum mechanics. In this context, the theory of decoherence constitutes the most accepted framework to bridge the gap between the classical and the quantum world \cite{breuer2002theory,zurek2003decoherence,schlosshauer2005decoherence,schlosshauer2007decoherence}. Classicality is here interpreted as an emergent property that arises as a quantum system coupled to an environment losses its coherence and becomes diagonal in its pointer states, which survive the harmful interaction. 

Beyond decoherence, the emergence of an objective reality can be interpreted through the concept of redundancy, assuming that during the interaction the information about the pointer states of the system is redundantly imprinted in the many degrees of freedom of the environment. Thereby, by measuring separate environmental fragments, independent observers can reach consensus about the classical state of the system without perturbing the pointer states, given that they are not subjected to direct measurements. The fact that some particular states of the system are the ones that survive among all the others and are able to procreate despite of the detrimental effects of its surrounding, is what gives this process the name of quantum darwinism (QD) \cite{ollivier2004objective, ollivier2005environment,blume2005simple, zurek2009quantum,paz2009redundancy, brandao2015generic,garcia2020decoherence,zurek2021emergence,touil2021eavesdropping}.            

In general, QD has been studied in the simplest scenario, where the fragments of the environment do not interact with each other. However, this may not be the most realistic situation in an experimental setup, where the intrinsic dynamics of the environment can play a significant role in the decoherent dynamics of the open system \cite{mirkin2021quantum, mirkin2021sensing}. For this reason, it is essential to understand how different environmental properties influence the emergence of QD and which are the mechanisms that may enhance the proliferation of redundancy. For instance, it has been shown that non-Markovianity hinders objectivity by suppressing the redundant records on the environment due to information backflows \cite{giorgi2015quantum, galve2016non, pleasance2017application,milazzo2019role}. Also, the degree of redundancy may be worsened by small changes in the microscopic description of the environment \cite{zwolak2009quantum,zwolak2010redundant,balaneskovic2015random, campbell2019collisional, ryan2020quantum, lorenzo2020anti}. More recently, a distinction between classicality and objectivity has been proposed for bipartite open quantum systems \cite{ccakmak2021quantum}. Finally, despite QD has been studied experimentally in photonic setups and nitrogen-vacancy centers \cite{ciampini2018experimental, chen2019emergence, unden2019revealing}, a full characterization taking into account the internal interactions of the environment remains unexplored. 

Based on the intuition that in the most realistic scenario a many-body environment should exhibit some degree of disorder as well as intra-environment interactions, our main goal in this work is to study how both of these environmental conditions influence the proliferation of redundancy. To do so, we use as an environmental model a disordered spin chain widely studied in the context of many-body localization (MBL) \cite{avishai2002level, santos2004integrability, vznidarivc2008many, pal2010many, de2013ergodicity, luitz2015many, solorzano2021multifractality}. This system exhibits an ergodic or a localized behaviour, depending on its energy and disorder strength, which sets a many-body mobility edge that has been estimated both theoretically and experimentally \cite{alet2018many,gong2021experimental,guo2021observation}. Thereby, by coupling a two-level quantum system to this disordered environment, we study the proliferation of redundant information both in the ergodic and localized phase. To this end, we introduce the notion of lack of redundancy and find that a high degree of disorder is hugely beneficial for the emergence of QD. This is related to the low entanglement that the eigenstates of the environment have in the localized regime, which enhances the capability of each fragment to store and retain information about the system. In fact, we show that the lack of redundancy exhibits the same scaling behaviour as the entanglement entropy of the initial state of the environment, which allows us to estimate the many-body mobility edge by means of our darwinistic measure. A remarkable implication of the latter is the existence of a critical degree of disorder beyond which the degree of objectivity increases the larger the environment is, evidencing the key role that disorder may play when the environment is of a thermodynamic size. At last, while previous works have shown that allowing a small interaction between the bath registers usually spoils the stored information in the environment \cite{riedel2012rise,giorgi2015quantum}, we find that a highly disordered evolution reduces this detrimental effect and enhances objectivity. 

This manuscript is organized as follows. In Section II we present the general framework, where we first introduce the physical system under consideration and then we present the main ideas behind the theory of QD. Also, we define the darwinistic measure that is used to quantify the degree of classical objectivity. In Section III, we begin by illustrating our darwinistic measure with a representative example and then we present our main results relating the degree of disorder, localization and QD. We conclude in Section IV with some final remarks.     

\section{General framework}
\subsection{Physical model}
To analyze the effect of disorder in the emergence of QD, we will consider a two-level quantum system $\mathcal{S}$ coupled to a disordered environment $E$. The total Hamiltonian describing the system plus environment is given by
\begin{equation}
    \hat{H}= \hat{H}_{int} + \lambda \hat{H}_{E},
\end{equation}
where $\hat{H}_{int}$ and $\hat{H}_{E}$ are the interaction and environmental Hamiltonian, respectively, and $\lambda$ is a parameter that regulates the influence of the intrinsic dynamics of $E$ with respect to the interaction with $\mathcal{S}$. In general, we remark that QD is studied in the limit where $\lambda \ll 1$, given that a small interaction between the bath registers usually spoils the stored information \cite{riedel2012rise,giorgi2015quantum}. Also, for simplicity, we will neglect the intrinsic Hamiltonian of $\mathcal{S}$ (assuming a dephasing interaction ($[\hat{H}_\mathcal{S}, \hat{H}_{int}]=0$),
the system time scales are not relevant and thus $\hat{H}_\mathcal{S}$ can be neglected). The open quantum system is coupled to the environment through a global interaction, given by
\begin{equation}
\hat{H}_{int}= \hat{\sigma}_z^{(\mathcal{S})} \otimes \sum_{k=1}^{L} \hat{\sigma}_y^{(k)},  
\end{equation}
where $\hat{\sigma}_{z}^{(\mathcal{S})}$ refers to the Pauli operator with direction $\hat{z}$ acting on $\mathcal{S}$, $\hat{\sigma}_{j}^{(k)}$ is the Pauli operator at site $k=\set{1,2,...,L}$ with direction $j=\set{x,y,z}$ and $L$ is the number of spins in the environment. The latter consists on a spin chain with nearest-neighbor interaction coupled to a random magnetic field in the $\hat{z}$ direction at each site. The environmental Hamiltonian is described by 

\begin{equation}
\hat{H}_E=\sum_{k=1}^{L}\left(\hat{S}_{k}^{x}\hat{S}_{k+1}^{x} + \hat{S}_{k}^{y}\hat{S}_{k+1}^{y} + \hat{S}_{k}^{z}\hat{S}_{k+1}^{z} \right) + \sum_{k=1}^{L} h_k^z \hat{S}_{k}^{z},     
\label{heisenberg}
\end{equation}
where $\{h_k^z\}$ is a set of random variables uniformly distributed within the interval $[-h,h]$ and periodic boundary conditions $\hat{S}^{x,y,z}_{1}=\hat{S}^{x,y,z}_{L+1}$ are considered. In this model, the $\hat{z}$ component of the total spin  $\hat{S}^z=\sum_{k=1}^L \hat{S}_k^z$ is a conserved quantity. This conservation allows the separation of the spanned space into smaller subspaces $\hat{\mathcal{S}}_n$, where $n$ is a fixed quantity of spins up or down. The dimension of each subspace is given by 
\begin{equation}
    D_n= \dim\left(\hat{\mathcal{S}}_n\right)= \begin{pmatrix}
L\\
n 
\end{pmatrix}
=\frac{L!}{n!(L-n)!}.
\end{equation}
We emphasize the fact that this system was widely studied in the context of MBL and the existence of a many-body mobility edge separating an ergodic and a localized phase has been demonstrated both theoretically and experimentally \cite{luitz2015many,alet2018many,gong2021experimental}. More specifically, while for weak disorder the system is chaotic and satisfies the eigenstate thermalization hypothesis (ETH), if the amount of disorder surpasses a certain critical threshold there is a transition to a MBL phase and the system do not thermalizes. 

Given that we are interested in the proliferation of redundancy and this is a dynamical phenomenon involving actively both the system and each individual fragment of the environment, first we will simulate the unitary dynamics of the entire system and then trace over the reduced part in which we need to focus. For this purpose, we will consider a separable initial state of the form
\begin{equation}
 \ket{\psi(0)}= \ket{+,x}_\mathcal{S}\otimes \ket{\xi}_{\epsilon},
 \label{initial_state}
\end{equation}
where $\ket{+,x}_\mathcal{S}$ is the eigenstate of $\hat{\sigma}_x^{(\mathcal{S})}$ with positive projection and $\ket{\xi}_{\epsilon}$ refers to the eigenstate of $\hat{H}_E$ closest to a normalized energy target $\epsilon$ defined as $\epsilon=(E - E_{min})/(E_{max}-E_{min})$. For simplicity, we will restrict our simulations to the subspace of zero magnetization $\hat{S}_0$ for even-sized chains and to the $\hat{S}_1$ sector for the odd ones. Consequently, given an initial state predetermined by a normalized energy target $\epsilon$ in the corresponding symmetric subspace, the procedure will consist on averaging over several disorder realizations for each fixed value of $h$. 

Considering the dephasing interaction between $\mathcal{S}$ and the environmental chain, the reduced density matrix of $\mathcal{S}$ can be solved analytically and is equal to

\begin{equation}
\begin{aligned}
\hat{\rho}_{\mathcal{S}}(t)  & = \frac{1}{2} \Big( \ket{0}\bra{0}
+ \ket{1}\bra{1} + r(t) \ket{0}\bra{1} +  r^{*}(t)\ket{1}\bra{0} \Big),
\end{aligned}
\label{rho_t}
\end{equation}
where $r(t)$ is known as the decoherence factor and in our situation is given by
\begin{equation}
    \begin{aligned}
    r(t)=\langle \xi_\epsilon| e^{i t\left[\lambda \hat{H}_{E}- \hat{H}_{\mathcal{S}\mathcal{E}} \right]} e^{-i t\left[\lambda \hat{H}_{E}+ \hat{H}_{\mathcal{S}\mathcal{E}} \right]}|\xi_\epsilon\rangle.
    \end{aligned}
\end{equation}
Let us note that $\hat{H}_{\mathcal{S}\mathcal{E}}$ refers to the term of $\hat{H}_{int}$ acting solely over the environmental degrees of freedom (in our case $\sum_{l=1}^{L}\hat{\sigma}_{l}^{y}$). In the traditional limit of QD, we can take as a first approximation the case where $\lambda \ll 1$, which simplifies the expression of the decoherence factor to
\begin{equation}
    \begin{aligned}
    r(t)  & \simeq \langle \xi_\epsilon| e^{-2i t \hat{H}_{\mathcal{S}\mathcal{E}}} |\xi_\epsilon\rangle \\
    &= \cos(2t) - i \sin(2t) \langle \xi_\epsilon| \hat{H}_{\mathcal{S}\mathcal{E}}
    |\xi_\epsilon\rangle.
    \end{aligned}
\end{equation}
Under this simple assumption, it is straightforward to calculate the purity of $\mathcal{S}$ as 
\begin{equation}
\begin{aligned}
\mathcal{P}_\mathcal{S}(t) &= \frac{1+r^2(t)}{2} \\
    & = \frac{1}{2}+\frac{\cos^2(2t)+\sin^2(2t) \langle \xi_\epsilon| \hat{H}_{\mathcal{S}\mathcal{E}}
    |\xi_\epsilon\rangle}{2}.
\end{aligned}
\end{equation}
Consequently, as far as the information content of the state of $\mathcal{S}$ is considered \cite{giorgi2015quantum}, the first revival occurs at $t=\pi/2$, where we have $\mathcal{P}_\mathcal{S}=1$. On the contrary, $t=\pi/4$ is the moment when the influence of $E$ over $\mathcal{S}$ is maximized before the first revival occurs. For this reason, we also expect QD to be more evident at this particular time.

\subsection{Quantum Darwinism}
The main ambition of QD is to elucidate how much information a fragment $\mathcal{F}$ of the environment acquires during the interaction with the system $\mathcal{S}$ and how redundant this information is.
In this framework, it is useful to focus on the mutual information between $\mathcal{S}$ and $\mathcal{F}$, which is defined as 
\begin{equation}
I(\mathcal{S}: \mathcal{F})=H_{\mathcal{S}}(t)+H_{\mathcal{F}}(t)-H_{\mathcal{S} \mathcal{F}}(t),    
\end{equation}
where $H_{\mathcal{S}}(t)$ and $H_{\mathcal{F}}(t)$ refer to the von Neumann entropies at time $t$ of $\mathcal{S}$ and $\mathcal{F}$, respectively, and $H_{\mathcal{SF}}(t)$ is the joint entropy between the two. In the case where $\mathcal{S}$ and $\mathcal{F}$ are initially uncorrelated, the mutual information quantifies the total information the fragment $\mathcal{F}$ gains about the state of the system $\mathcal{S}$. In order to quantify the degree of redundancy achieved during the interaction, we will define the notion of perfect redundancy as the ideal case where the mutual information between $\mathcal{S}$ and $\mathcal{F}$ is equal to the entropy of $\mathcal{S}$ for any possible fraction $f$ of the environment considered, i.e. $\mathcal{I}(\mathcal{S}:\mathcal{F}_l)=H_\mathcal{S} \, \forall \, f=l/L<1$, where $\mathcal{F}_l$ is an environmental fragment composed of $l$ different components (spins) and $L$ is the total number of spins of the entire environment. Consistently, in the case where the fragment is the whole environment ($f=1$), then $H_{\mathcal{SF}}=0$ and we have $\mathcal{I}(\mathcal{S}:\mathcal{F}_L)=2H_\mathcal{S}$. It is important to notice that to avoid any possible bias when selecting the environmental fragment, we must compute an averaged mutual information $\overline{\mathcal{I}}(\mathcal{S}:\mathcal{F}_l)$, where the average is taken over all possible ways of composing the fragment of $l$ components given the environment of $L$ spins (there are $\frac{L!}{l!(L-l)!}$ possible combinations). With the definitions above, it is straightforward to define the lack of redundancy ($\mathcal{LR}$) as the difference between perfect redundancy and the averaged mutual information achieved during the dynamics, i.e.
\begin{equation}
\mathcal{LR}=\sum\limits_{l=1}^{L-1} \frac{|H_\mathcal{S} - \overline{\mathcal{I}}(\mathcal{S}:\mathcal{F}_l)|}{H_\mathcal{S}}.    
\end{equation}
This constitutes the measure of objectivity that we will use from now on. We remark that different measures have also been previously used in the literature to quantify redundancy \cite{zwolak2009quantum,zwolak2010redundant}, but the one proposed in this work is more reliable for comparing environments with slightly different sizes. This is due to the fact that given the fractional nature of the fragment size $(f=l/L)$, the possible set of values that $f$ can take is different depending on which is the particular size of the chain.  

\section{Darwinism, MBL and interactions}
\begin{figure}[!htb]
\begin{center}
\includegraphics[width=87mm]{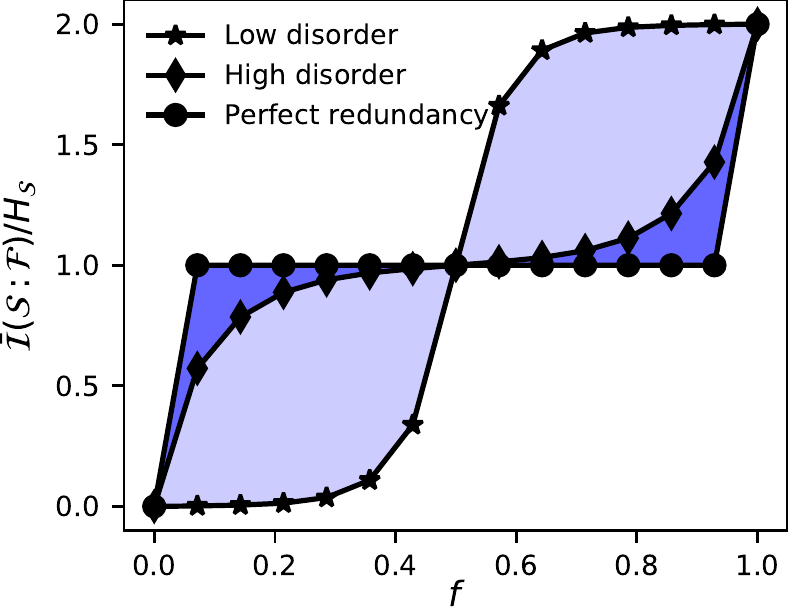}
\begin{footnotesize}
\caption{Rescaled averaged mutual information $\overline{\mathcal{I}}(\mathcal{S}:\mathcal{F})/H_\mathcal{S}$ as a function of the environmental fragment size $f$ for an environment composed of $L=14$ spins. Given a fixed environmental size and disorder strength, the filled area between perfect redundancy and the averaged mutual information obtained constitutes the lack of redundancy ($\mathcal{LR}$) and our measure of objectivity. The parameters are set as $\epsilon=0.5$, $t=\pi/4$, $\lambda=0$, $h=0.01$ for low disorder and $h=5.0$ for high disorder. In both situations, 1000 realizations for different sets of $\set{h_k^z}\in[-h,h]$ were considered.}
\label{redundancy}
\end{footnotesize}
\end{center}
\end{figure}

In this Section, we present our main results that shed light on the relationship between disorder and QD. As a first illustrative example to clarify the definitions of the previous Section, we start by plotting in Fig. \ref{redundancy} the rescaled averaged mutual information $\overline{\mathcal{I}}(\mathcal{S}:\mathcal{F}_l)/H_\mathcal{S}$ as a function of the size of the environmental fragment for an environment composed of $L=14$ spins. Additionally, in the same plot we show the averaged lack of redundancy $\mathcal{LR}$ as a blue light (dark) filled area in the case with low (high) disorder. It is important to remark that in this simulation the intrinsic dynamics of the environment is completely neglected in the evolution by setting $\lambda=0$. Hence, disorder only comes into play in the particular eigenstate of the environment that is taken as an initial state (see Eq. (\ref{initial_state})). 

\renewcommand{\figurename}{Figure} 

From Fig. \ref{redundancy} we can notice that the presence of high disorder seems to enhance the emergence of objectivity, leading to a better plateau in the averaged mutual information shared between the system and the environment. Beyond this qualitative result, it is well-known that in disordered spin chains, such as our environmental model, the degree of disorder is strictly related to the localization of the system. In particular, if the disorder is weak enough the system is ergodic and its eigenstates are highly entangled. On the contrary, as the degree of disorder increases, the system reaches a localized phase and its eigenstates exhibit much less entanglement. Taking into account this fact, in what follows we will delve into the qualitative result obtained in Fig. 1. To do so, in the following subsection we will perform a systematic analysis exploring the role 
that the localization in the initial state of the environment plays in its capability of enabling redundancy.

\subsection{Localization in the initial state}
\begin{figure}[!htb]
\begin{center}
\includegraphics[width=87mm]{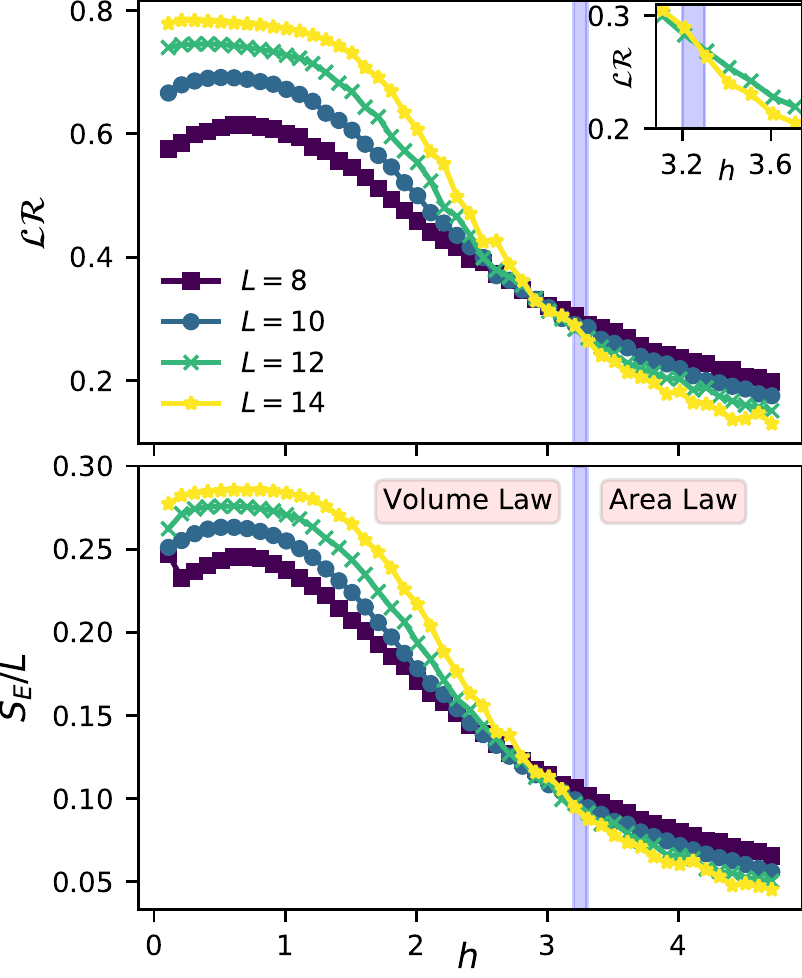}
\begin{footnotesize}
\caption{\textit{Upper panel:} Lack of redundancy $\mathcal{LR}$ for different environmental sizes as a function of the disorder strength $h$. \textit{Inset:} Zoom into the transition of $\mathcal{LR}$, signaled by the intersection between the two curves with the largest environmental size. The parameters are set as $\epsilon=0.5$, $t=\pi/4$ and $\lambda=0$. \textit{Lower panel:} Entanglement entropy per site $S_E/L$ of a single eigenstate of $\hat{H}_E$ ($\ket{\xi}_{\epsilon=0.5}$) for different environmental sizes as a function of the disorder strength $h$. A crossover between a volume and area law is observed for a critical disorder $h_c\simeq3.2$ (blue separation). In both panels, all curves are averaged over at least 1000 different realizations of disorder.}
\label{redundancy2}
\end{footnotesize}
\end{center}
\end{figure}

To further explore how the localization in the initial state of the environment influences the emergence of QD, we will now focus on the half-chain entanglement entropy  $S_E =-\Tr_A \left( \hat{\rho}_A \ln \hat{\rho}_A \right)$ of the reduced density matrix of the environment $\hat{\rho}_A=\Tr_B \ket{\xi}\bra{\xi}$, where the traces are over the left and right half-chain Hilbert spaces, respectively, and $\ket{\xi}$ is an eigenstate of $\hat{H}_E$. In the localized regime, the reduced density matrix $\hat{\rho}_A$ of a typical eigenstate $\ket{\xi}$ possesses low entanglement entropy and an area-law scaling. Conversely, in a chaotic regime satisfying ETH, eigenstates are highly entangled and exhibit a volume-law scaling. Thereby, it is possible to distinguish both regimes by analyzing the scaling behaviour of $S_E$. To do so, in Fig. \ref{redundancy2} we plot the entanglement entropy per site $S_E/L$ (lower panel), together with the lack of redundancy $\mathcal{LR}$ (upper panel), as a function of the disorder strength $h$ for different environmental sizes. As before, we restrict ourselves to the zero magnetization sector and entirely focus on the eigenstate $\ket{\xi}_\epsilon$, i.e. the one with energy closest to the normalized target $\epsilon$ in each disorder realization. Once again, for computing $\mathcal{LR}$, we neglect the influence of the internal dynamics of the environment in the evolution by setting $\lambda=0$.

\renewcommand{\figurename}{Figure} 

The first conclusion we can extract from Fig. \ref{redundancy2} is that both $\mathcal{LR}$ and $S_E/L$ exhibit a similar behaviour as a function of disorder. In particular, we can observe that if the disorder strength $h$ is lower than a critical value $h_c$, both $\mathcal{LR}$ and $S_E/L$ increase with increasing $L$. On the contrary, if $h>h_c$ the behaviour of both quantities is the opposite. What does this mean in terms of QD? Interestingly, this means that there is a critical value of disorder beyond which the degree of objectivity rises the larger the environment is. We emphasize that this critical value of $h_c$ has been estimated by looking at the intersection between the two curves of largest dimension (see inset in the upper panel of Fig. \ref{redundancy2}). 

\begin{figure}[!htb]
\begin{center}
\includegraphics[width=87mm]{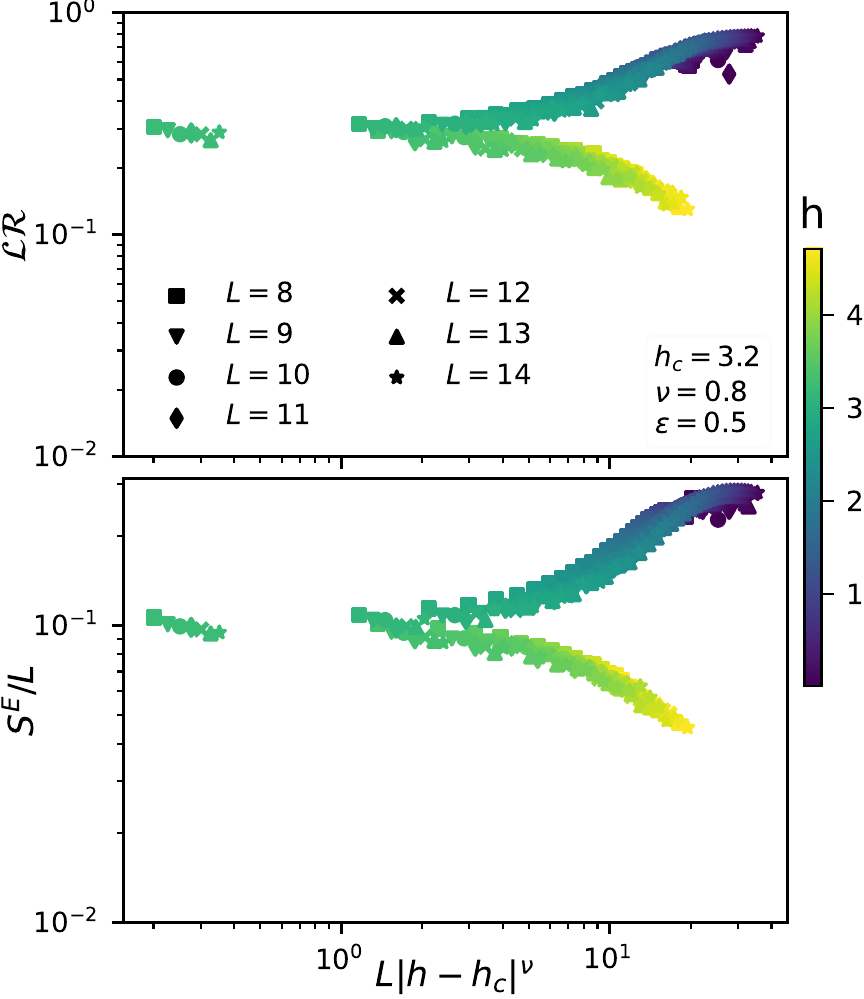}
\begin{footnotesize}
\caption{\textit{Upper panel:} Lack of redundancy $\mathcal{LR}$ as a function of $L|h-h_c|^\nu$. The parameters are set as $\epsilon=0.5$, $t=\pi/4$ and $\lambda=0$. \textit{Lower panel:} Entanglement entropy per site $S_E/L$ of a single eigenstate of $\hat{H}_E$ ($\ket{\xi}_{\epsilon=0.5}$) as a function of $L|h-h_c|^\nu$. 
For weak disorder, there is a volume-law scaling that leads to a constant $S_E/L$, while for strong disorder we have an area law characterized by a decreasing $S_E/L$.
For the odd-sized chains, only the subspace $\hat{S}_1$ was considered. All curves are averaged over at least 1000 different realizations of disorder.}
\label{redundancy3}
\end{footnotesize}
\end{center}
\end{figure}

To provide further insight about this transition, we perform a finite size scaling analysis by collapsing all the data to the form $g[L^{1/\nu}(h-h_c)]$, similarly to what was done by previous works dealing with the same disordered quantum system  \cite{luitz2015many,alet2018many}. The results of the scaling are shown in Fig. \ref{redundancy3}. In particular, in the lower panel we can see that the transition is characterized by a change in the entanglement entropy scaling from an area law for $h>h_c$ (where $S_E/L \to 0$) to a volume law for $h<h_c$ (where $S_E/L \to \text{constant}$). As was claimed before, the same scaling behaviour is observed for the lack of redundancy $\mathcal{LR}$, as we show in the upper panel of Fig. \ref{redundancy3}.   

\renewcommand{\figurename}{Figure} 

\begin{figure}[!htb]
\begin{center}
\includegraphics[width=87mm]{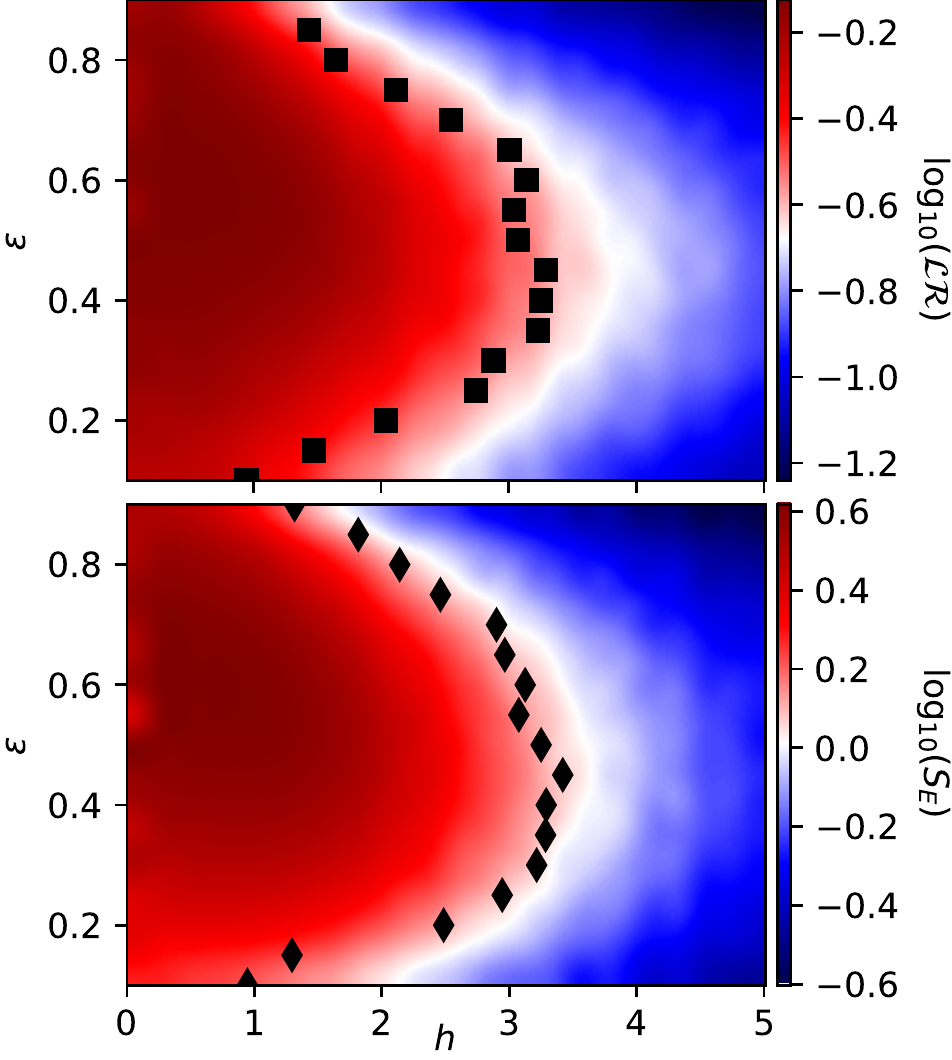}
\begin{footnotesize}
\caption{\textit{Upper panel:} Lack of redundancy $\mathcal{LR}$ as a function of the disorder strength $h$ for different eigenstates with energy $\epsilon$ taken as initial states. Since this quantity is much harder to calculate, a smaller environment of $L=12$ was considered. The black squares refer to the critical values of $h_c$, that were estimated by looking at the intersection between the data obtained for $L=12$ and $L=10$, respectively.
Parameters are set as $t=\pi/4$, $\lambda=0$.
\textit{Lower panel:} Entanglement entropy $S_E$ as a function of the disorder strength $h$ and for different eigenstates of $\hat{H}_E$ with normalized energies $\epsilon$ (i.e. $\ket{\xi}_{\epsilon}$). An environment composed of $L=14$ was considered in this case. The black diamonds refer to the critical values of $h_c$, that were estimated by looking at the intersection between the data obtained for $L=14$ and $L=12$, respectively. Both panels are averaged over 1000 different realizations of disorder.}
\label{comparison}
\end{footnotesize}
\end{center}
\end{figure}

Despite in this first part we have shown a close relationship between the entanglement of the initial state of the environment and its capability of enabling redundancy, it is important to notice that until now we have restricted entirely to the middle of the spectrum by considering the eigenstate with energy closest to $\epsilon=0.5$ in all simulations. However, the value of the critical disorder $h_c$ depends on the energy under consideration, which determines what is called a many-body mobility edge \cite{vznidarivc2008many, de2013ergodicity,luitz2015many, alet2018many,guo2021observation, gong2021experimental,kjall2014many,pal2010many,solorzano2021multifractality,zhang2021mobility}. For this reason, it is worthy to study how our darwinistic measure $\mathcal{LR}$ behaves when considering different eigenstates of $\hat{H}_E$ as initial states. This is precisely what is shown in Fig. \ref{comparison}, where we compare our measure of objectivity $\mathcal{LR}$ with the entanglement entropy $S_E$ in the same region of parameters set both by $h$ and $\epsilon$. 

\renewcommand{\figurename}{Figure} 

Once again, it is clear that both quantities exhibit almost the same structure, which means that a low $S_E$ in the initial state of the environment implies a better capacity to store redundant information and thus greater objectivity. Thereby, taking into account that the region of low entanglement is always linked to a high degree of disorder, we can conclude that disorder is beneficial for the emergence of QD. As a complement, in both panels of Fig. \ref{comparison} we have estimated the many-body mobility edge with the same procedure followed before for $\epsilon=0.5$, finding a very good agreement between both measures. This mobility edge is shown with black squares for $\mathcal{LR}$ and with black diamonds for $S_E$. 

After all the simulations shown in this subsection, we can conjecture that the reason why objectivity is boosted by using an eigenstate of a highly disordered environment as initial state, is strictly related to the low entanglement that eigenstates have in the MBL phase. This low entanglement enables each environmental fragment to store more information about the system since its initial state is much closer to a pure state. On the contrary, if the disorder is too low, eigenstates are highly entangled due to the ergodic nature of the system and consequently the initial state of each fragment is much nearer to a maximally mixed state. In this situation, the storing of information is shrinked and redundancy cannot be achieved.

\subsection{Influence of intra-environment interaction}
\begin{figure}[!htb]
\begin{center}
\includegraphics[width=87mm]{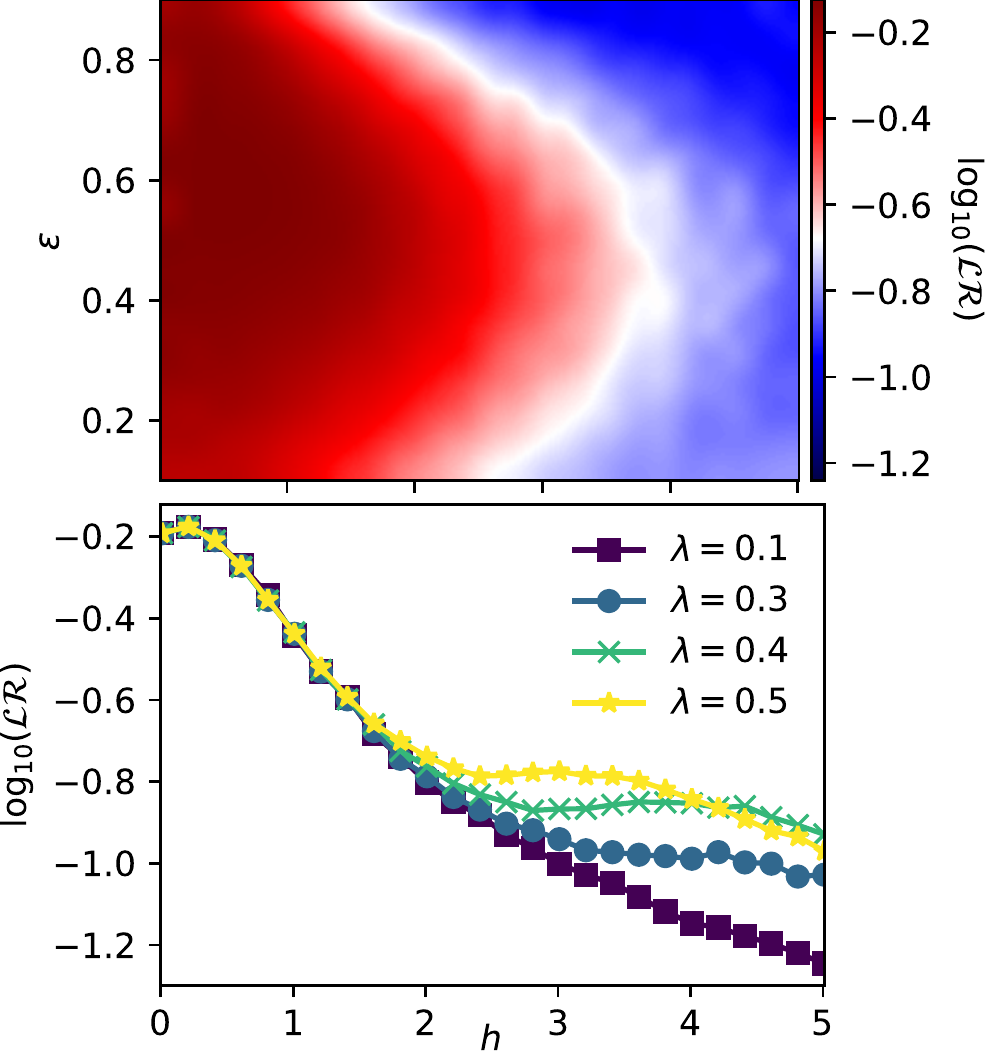}
\begin{footnotesize}
\caption{\textit{Upper panel:} Lack of redundancy $\mathcal{LR}$ as a function of the disorder strength $h$ and using as initial states different eigenstates with energy $\epsilon$. In this panel, parameters are set as $L=12$, $t=\pi/4$ and $\lambda=0.3$. \textit{Lower panel:} Lack of redundancy $\mathcal{LR}$ as a function of the disorder strength $h$ for different values of $\lambda$. In this panel, parameters are set as $L=12$, $t=\pi/4$ and $\epsilon=0.9$. In both panels, 1000 different realizations of disorder were considered.}
\label{dynamics}
\end{footnotesize}
\end{center}
\end{figure}

A key point that we have not analyzed yet is the following: what if we allow a mixing of the bath records by setting $\lambda \neq 0$ in the evolution? Based on our previous analysis, we already know that the initial states of the environment belonging to the ergodic regime are incapable of storing redundant information and thus of enabling objectivity. On the contrary, how robust are the MBL initial states to the mixing due to the internal dynamics of the environment? Is it still possible to distinguish both regimes in the presence of this effect? To address these questions, we will first take as initial state an eigenstate of the environment with a given disorder $h$ and energy $\epsilon$, evolve the system with the same amount of disorder (i.e. $\hat{H}_{int} + \lambda \hat{H}_{E}|_h$) and finally average over several realizations. The results of this analysis are shown in Fig. \ref{dynamics}, where we plot $\mathcal{LR}$ for different values of $\lambda$ (the parameter that regulates the influence of the intra-environment interactions in the evolution). 

For instance, in the upper panel of Fig. \ref{dynamics}, where we set $\lambda=0.3$, we can observe that the same structure obtained before for $\lambda=0$ holds. Despite the degree of objectivity is slightly worsened in this case for the MBL region, we can still distinguish the ergodic from the localized regime by means of our darwinistic measure. Additionally, in the lower panel of Fig. \ref{dynamics} we show how the lack of redundancy behaves for different values of $\lambda$ as a function of the disorder strength $h$. It is clear from here that for weak disorder the intra-environment interactions have no influence at all in the redundancy. This is not surprising given that even in the case of $\lambda=0$ objectivity was not possible in this region. On the contrary, as the amount of disorder increases and we approach the MBL regime, the influence of the intra-environment dynamics is more notorious and we can appreciate how objectivity is slightly spoiled as $\lambda$ increases. 

\begin{figure}[!htb]
\begin{center}
\includegraphics{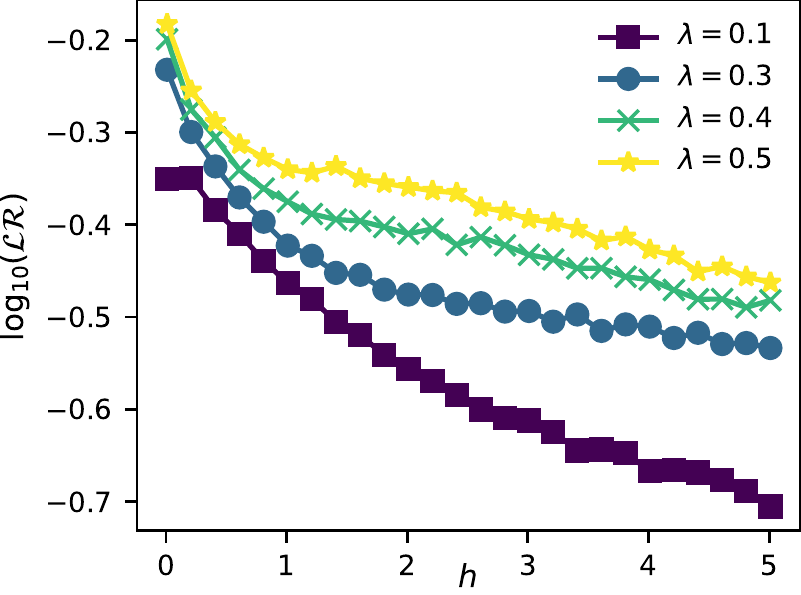}
\begin{footnotesize}
\caption{Lack of redundancy $\mathcal{LR}$ as a function of the disorder strength $h$ for different values of $\lambda$. Parameters are set as $L=10$, $t=\pi$ and the initial state of the environment is always fixed as $\ket{\xi}\mid_{h=5.0,\epsilon=0.5}$. As usual, 1000 different realizations of disorder were considered.}
\label{dynamics2}
\end{footnotesize}
\end{center}
\end{figure}

Finally, let us now restrict ourselves to the most favorable situation by fixing the initial state of the environment as a MBL eigenstate. Unlike our previous analysis, we will now evolve the same highly localized initial state with different amounts of disorder. Is the internal dynamics of a highly disordered environment still advantageous in this situation? To address this important question, in Fig. \ref{dynamics2} we show the results of a simulation where we fix the initial state of the environment as $\ket{\xi}\mid_{h=5.0,\epsilon=0.5}$ and then evolve the entire system considering different values of disorder strength, ($\hat{H}_{int}+\lambda \hat{H}_E \mid_{h'}$). From this simulation we can conclude that the presence of disorder is still beneficial for the appearance of objectivity, even in the presence of intra-environment interactions and independently of the initial state under consideration.    

\section{Conclusions}
The theory of QD studies how classical objectivity emerges from a quantum realm. It poses that besides of washing out the coherences of the open quantum system to which it is coupled, the environment also acts as an active witness by redundantly storing the information about the system's state. Hence, independent observers can reach consensus on the actual state of the system by measuring separate environmental fragments. Based on the intuition that in a realistic scenario a many-body environment should exhibit some degree of disorder, in this work we have explored the role that this disorder plays in the emergence of QD.

By modelling the environment as a 1D disordered spin chain, we have found that a high degree of disorder is greatly beneficial for the emergence of classical objectivity. In particular, we have shown that this is a consequence of the low entanglement that the eigenstates of the environment, which are taken as the initial states, exhibit in the localized regime. This enables each individual fragment to store a greater amount of redundant information, in comparison to the highly entangled environmental eigenstates that belong to the ergodic regime at low disorder. In particular, we have shown that the EE of the initial state not only exhibits the same structure as the lack of redundancy but also shares the same scaling behaviour. This fact allowed us to estimate the many-body mobility edge by means of our darwinistic measure, yielding a consistent agreement in comparison to the EE. In regards to QD, this result means that if the degree of disorder is higher (lower) than a critical value, the redundancy increases (decreases) the larger the environment is. Therefore, our findings evidence the important role that disorder can play in a realistic situation where the environment is of a thermodynamic size.

Also, we have analyzed the mixing of the redundant records by allowing 
intra-environment interactions. Despite low disorder is associated to an ergodic dynamics and ergodicity usually yields to a markovian evolution \cite{mirkin2021sensing}, we have found that a highly disordered evolution is less harmful for the appearance of classical objectivity. Hence, a high amount of disorder is not only beneficial for the initial state under consideration but also reduces the mixing of the redundant records in the presence of intra-environment interactions. At last, we sincerely hope our findings to shed new light on how classical objectivity emerges from a quantum world and we look forward to the possibility of implementing our ideas on some of the experimental platforms recently used in the study of the MBL transition \cite{guo2021observation}. 

\begin{acknowledgements}
The work was partially supported by CONICET (PIP 112201 50100493CO), UBACyT (20020130100406BA), and ANPCyT (PICT-2016-1056). 
\end{acknowledgements}



\bibliography{main.bib}
\end{document}